\documentclass[10pt,twocolumn]{article}

\usepackage[utf8]{inputenc}
\usepackage{amsmath}
\usepackage{amssymb}
\usepackage{amsthm}
\usepackage{titlesec}
\usepackage{cite}
\usepackage{booktabs}
\usepackage{graphicx}
\usepackage[colorlinks=false]{hyperref}
\usepackage[format=plain,labelfont=it]{caption}
\usepackage[left=1.5cm,right=1.5cm,top=2cm,bottom=2cm]{geometry}
\usepackage[noend]{algpseudocode}
\usepackage[FIGTOPCAP]{subfigure}

\usepackage{tikz}
\usetikzlibrary{arrows}
\usetikzlibrary{shapes}
\usetikzlibrary{decorations.pathmorphing}
\usetikzlibrary{decorations.pathreplacing}
\usetikzlibrary{decorations.shapes}
\usetikzlibrary{decorations.text}
\usetikzlibrary{positioning, shapes}

\titleformat{\section}{\centering\normalfont\scshape}{\Roman{section}.}{5pt}{}
\titleformat{\subsection}{\normalfont\it}{\Alph{subsection}.}{5pt}{}
\titleformat{\subsubsection}{\normalfont\it}{\hspace{4mm}\arabic{subsubsection})}{5pt}{}

\newcommand\infoFootnote[1]{%
  \begingroup
  \renewcommand\thefootnote{}\footnote{#1}%
  \addtocounter{footnote}{-1}%
  \endgroup}

\newcommand{\R}{\mathbb{R}} 
\newcommand{\N}{\mathbb{N}}

\newcommand{\Rc}{\mathcal{R}}

\newcommand{\lb}{\boldsymbol{l}}

\newcommand{\eb}{\boldsymbol{e}}
\newcommand{\yb}{\boldsymbol{y}}
\newcommand{\zb}{\boldsymbol{z}}

\newcommand{\xib}{\boldsymbol{\xi}}
\newcommand{\xb}{\boldsymbol{x}}

\newcommand{\bb}{\boldsymbol{b}}

\newcommand{\fb}{\boldsymbol{f}}
\newcommand{\Wb}{\boldsymbol{W}}

\newcommand{\Gb}{\boldsymbol{G}}
\newcommand{\Qb}{\boldsymbol{Q}}

\newcommand{\Pb}{\boldsymbol{P}}

\newcommand{\gb}{\boldsymbol{g}}

\newcommand{\zerob}{\boldsymbol{0}}

\newcommand{\Phib}{\boldsymbol{\Phi}}

\newcommand{\norm}[1]{\left\lVert#1\right\rVert}

\newcommand{\blind}[1]{\textcolor{white}{#1}}

\newtheorem{thm}{Theorem}

\newtheorem{rem}{Remark}
\newtheorem{cor}[thm]{Corollary}

\newtheorem{conj}{Conjecture}
\newtheorem{alg}{Algorithm}

\title{\vspace{-2mm}\bf Tailored max-out networks for learning convex PWQ functions}
\author{Dieter Teichrib and Moritz Schulze Darup\vspace{2mm}}
\date{}

\begin{document}

\maketitle

\textbf{\textit{Abstract}.} {\bf Convex piecewise quadratic (PWQ) functions frequently appear in control and elsewhere. %often appear in the field of machine learning. 
For instance, it is well-known that the optimal value function (OVF) as well as Q-functions for linear MPC are convex PWQ functions.
Now, in learning-based control, these functions are often represented with the help of artificial neural networks (NN).
In this context, a recurring question is how to choose the topology of the NN in terms of depth, width, and activations in order to enable efficient learning.
An elegant answer to that question could be a topology that, in principle, allows to exactly describe the function to be learned.
Such solutions are already available for related problems. In fact, suitable topologies are known for piecewise affine (PWA) functions that can, for example, reflect the optimal control law in linear MPC. Following this direction, we show in this paper that convex PWQ functions can be exactly described by  max-out-NN with only one hidden layer and two neurons.}
% leave no space here
\infoFootnote{D. Teichrib and M. Schulze Darup are with the \href{https://rcs.mb.tu-dortmund.de/}{Control and~Cyber-physical Systems Group}, Faculty of Mechanical Engineering, TU Dortmund University, Germany. E-mails:  \href{mailto:moritz.schulzedarup@tu-dortmund.de}{\{dieter.teichrib, moritz.schulzedarup\}@tu-dortmund.de}. \vspace{0.5mm}}
\infoFootnote{\hspace{-1.5mm}$^\ast$This paper is a \textbf{preprint} of a contribution to the 20th European Control Conference 2022.}

\section{Introduction}

Learning-based and data-driven methods are heavily used in all kind of sectors including health, finance, transportation, industry, or energy.
While the applications are as diverse as the sectors, using supervised or reinforcement learning in order to approximate an unknown function is still a recurring task. 
Popular choices for the models to be learned then are, for example, artificial neural networks (NN) or Bayesian networks. In both cases, choosing a suitable network topology  is crucial for efficient learning. As well-founded design rules are rare, choices are often made by trial-and-error. However, some tasks allow for more educated guesses. For instance, if the function to be approximated is known to be piecewise affine (PWA), NN with max-out or rectified linear unit (ReLU) activations are ideal as their input-output relation is likewise PWA. Moreover, if also the number of affine segments is available, then it is even possible to specify NN that, in principle, allow to exactly describe the function of interest (see, e.g., \cite{Montufar2014,Arora2016,Ferlez2020}). This observation has led to tailored NN for (linear)  model predictive control (MPC) \cite{Chen2018,Karg2020,SchulzeDarup_2020_ECC_ANN}, where the optimal control law is well-known to be PWA \cite{Bemporad2002}.

Another interesting property of linear MPC is that the optimal value function (OVF) is piecewise quadratic (PWQ) and convex. In this context, the OVF or related Q-functions play a central role for data-driven predictive control based on reinforcement learning \cite{Markolf2021,Zhong2013}. Nevertheless, tailored NN for representing convex PWQ have only been rarely considered. One exception can be found in \cite{Teichrib2021}, where we recently showed that one-dimensional PWQ functions with $s$ segments can be exactly described by a ReLU-NN with one hidden layer of width $2s$ (and an augmented input). While this observation was interesting, the underlying method is restricted in that an extension to the multi-dimensional inputs (or states in the context of MPC) seems hard if not impossible.
As a consequence, we reconsider the problem of identifying tailored NN for convex PWQ functions here. However, in contrast to \cite{Teichrib2021}, we will focus on max-out-NN as they seem more suitable for exploiting convexity and for addressing multi-dimensional inputs. In fact, it is well-known that convex PWA functions (with one or more inputs) can be trivially represented by one max-out neuron (per output). 
While this feature does, in general, not apply to convex PWQ functions $\varphi$, we will show that a systematic modification of $\varphi$ allows to derive a representation via two max-out neurons.

We organize the presentation of the novel approach as follows.
In Section~\ref{sec:background} we provide some preliminaries on PWQ functions and background on NN. Further, we briefly summarize the results from \cite{Teichrib2021}. We present our main result, i.e., tailored max-out-NN for exactly describing PWQ functions, in Section~\ref{sec:tailoredNN}. More specifically, we construct max-out-NN that exactly describes a convex PWQ function for the special case of one-dimensional inputs (i.e., $n=1$) and show a possible direction for an extension to functions with higher-dimensional inputs (i.e. $n>1$) on an exemplary function in the Section~\ref{sec:example}. Finally, in the aforementioned section we also illustrate our main results and state conclusions in Section~\ref{sec:conclusion}.

\section{Preliminaries and background}\label{sec:background}

Throughout the paper, we deal with convex PWQ functions of the form 
\begin{equation}\label{eq:PWQ}
\varphi(\xb) = \left\{
\begin{array}{cc}
\!\!\xb^\top\Qb^{(1)}\xb +\xb^\top \lb^{(1)} +c^{(1)} & \text{if}\,\,\,\xb\in\Rc^{(1)}, \\
\vdots &  \vdots \\
\!\!\xb^\top\Qb^{(s)}\xb +\xb^\top \lb^{(s)} +c^{(s)} & \text{if}\,\,\,\xb\in\Rc^{(s)}, \\
\end{array}
\right.\!
\end{equation}
for scalar or multi-dimensional inputs $\xb\in \R^n$. The parameters $\Qb^{(i)} \in \R^{n\times n}$,  $\lb^{(i)} \in \R^{(n\times 1)}$, and $c^{(i)}\in\R$ reflect the quadratic, the linear, and the constant parts of the various segments. The regions $\Rc^{(i)}\subset \R^n$ are convex polytopes with pairwise disjoint interiors.

\subsection{Neural networks with rectifier and max-out activations}

In general, a feed-forward-NN with $\ell \in\N$ hidden layers and $w_i$ neurons in layer $i$ can be written as a composition of the form
\begin{equation}\label{eq:ANN}
\Phib(\xib)=\fb^{(\ell+1)}\circ \gb^{(\ell)}\circ \fb^{(\ell)}\circ \dots \circ \gb^{(1)}\circ \fb^{(1)}(\xib).
\end{equation}
Here, the functions $\fb^{(i)}: \R^{w_{i-1}} \rightarrow \R^{p_i w_i}$ for $i \in \{1,\dots,\ell\}$ refer to preactivations, where the parameter $p_i\in \N$ allows to consider ``multi-channel'' preactivations as required for max-out. Moreover, $\gb^{(i)}: \R^{p_i w_i} \rightarrow \R^{w_i}$ stand for activation functions and $\fb^{(\ell+1)}: \R^{w_{\ell}} \rightarrow \R^{w_{\ell+1}}$ reflects  postactivation.
The functions $\fb^{(i)}$ are typically affine, i.e.,
\begin{equation}
\nonumber
\fb^{(i)}(\yb^{(i-1)})=\Wb^{(i)}\yb^{(i-1)}+\bb^{(i)},
\end{equation}
where $\Wb^{(i)}\in\R^{p_i w_i\times w_{i-1}}$ is a weighting matrix, $\bb^{(i)} \in \R^{p_i w_i}$ is a bias vector, and 
 $\yb^{(i-1)}$ denotes the output of the previous layer with $\yb^{(0)}:=\xib$. 

Now, various activation functions have been proposed. As already stated in the introduction, we here focus on ReLU and max-out neurons \cite{Goodfellow2013}. The corresponding activation functions are specified as
\begin{equation}
\nonumber
\gb^{(i)}(\zb^{(i)})=\max\left\{\zerob,\zb^{(i)}\right\}:=\begin{pmatrix}
\max\big\{0,\zb_1^{(i)}\big\} \\
\vdots \\
\max\big\{0,\zb_{w_i}^{(i)}\big\}
\end{pmatrix}
\end{equation}
for ReLU and
\begin{equation}
\nonumber
\gb^{(i)}(\zb^{(i)})=\begin{pmatrix}
\max \limits_{1\leq j \leq p_i}\big\{\zb_j^{(i)}\big\} \\
\vdots \\
\max \limits_{p_i(w_i-1)+1 \leq j \leq p_i w_i}\big\{\zb_{j}^{(i)}\big\}
\end{pmatrix}
\end{equation}
for max-out, where $\zb_j^{(i)}$ denotes the $j$-th component of $\zb^{(i)}\in \R^{p_i w_i}$ and where we use the shorthand notation
$$
\max\limits_{1\leq j \leq p_i}\big\{\zb_j^{(i)}\big\}:=\max\big\{\zb_1^{(i)},\dots,\zb_{p_i}^{(i)}\big\}.
$$
We will refer to the resulting networks as ReLU-NN and max-out-NN, respectively. 

\subsection{Tailored ReLU-NN for representing PWQ functions}
\label{subsec:tailoredReLU}

We recently showed in \cite{Teichrib2021} that PWQ functions can be represented by a ReLU-NN for the special case of scalar variables $\xb$ (i.e., $n=1$).
Since this case will also be in the focus of this paper, we specify the notation of~\eqref{eq:PWQ} for ease of presentation. Hence, we consider
\begin{equation}\label{eq:PWQ1D}
\varphi(x) = \left\{
\begin{array}{cc}
q_1 x^2 +l_1 x +c_1 & \text{if}\,\,\,x\in [\underline{x}_1,\overline{x}_1], \\
\vdots &  \vdots \\
q_s x^2 +l_s x +c_s  & \text{if}\,\,\,x\in[\underline{x}_s,\overline{x}_s] \\
\end{array}
\right.\!
\end{equation}
instead of \eqref{eq:PWQ} for the majority of the paper, where we assume that the relation $\underline{x}_i<\overline{x}_i=\underline{x}_{i+1}<\overline{x}_{i+1}$ holds for each $i \in \{1,\dots, s-1\}$ and that the interval $[\underline{x}_1,\overline{x}_s]$ is bounded. We further define
\begin{equation*}
\varphi_i(x):=q_i x^2 +l_i x +c_i.
\end{equation*}
The specification further allows to particularize (conditions for) continuity and convexity. In fact, these properties require  
\begin{subequations}
\label{eq:convAndContPhi}
\begin{align}
q_i \overline{x}_i^2 +l_i \overline{x}_i +c_i &= q_{i+1} \overline{x}_i^2 +l_{i+1} \overline{x}_i +c_{i+1},\\  % only x her\\
2 q_i \overline{x}_i+l_i  & \leq 2 q_{i+1} \overline{x}_i +l_{i+1} 
\end{align}
\end{subequations}
for every $i\in\{1,\dots,s-1\}$ as well as $q_i\geq 0$ for every $i\in\{1,\dots,s\}$.
Now, according to \cite[Cor.~3]{Teichrib2021}, \eqref{eq:PWQ1D} can be represented by a ReLU-NN with one hidden layer of width $w_1=2s$. In fact, \cite[Thm.~1]{Teichrib2021} combined with \cite[Thm.~1]{SchulzeDarup_2020_ECC_ANN} leads to the ReLU-NN 
\begin{equation}
\label{eq:ReLUoneLayer}
   \Phib(\xib)=\Wb^{(2)} \max\left\{ \zerob, \Wb^{(1)}\xib + \bb^{(1)} \right\} + \bb^{(2)} 
\end{equation}
with the parameters
\begin{align*}
    \Wb^{(1)}\!&:=\begin{pmatrix} 
\underline{x}_1+\overline{x}_1 & -1 \\ 
\vdots & \vdots \\ 
\underline{x}_s+\overline{x}_s & -1 \\ 
-1 & 0\\ 
\blind{+}1 & 0 \\ 
%\blind{+}1 & 0 \\ 
\blind{+}\vdots & \vdots \\ 
\blind{+}1 & 0 \\ 
\end{pmatrix}, \qquad \bb^{(1)}\!:=\begin{pmatrix}
-\underline{x}_1 \overline{x}_1 \\
\vdots \\
-\underline{x}_s \overline{x}_s \\
\blind{+}\overline{x}_1 \\
-\overline{x}_1 \\
%-\overline{x}_2 \\
\vdots \\
-\overline{x}_{s-1}
\end{pmatrix}, \\
\Wb^{(2)}\!&:=\begin{pmatrix} 
-q_1 & \!\!\dots\!\!\! & -q_s & \kappa_1 & \kappa_2 & \kappa_3\!-\!\kappa_2 & \!\!\dots\!\!\! & \kappa_s \!-\! \kappa_{s-1}
\end{pmatrix}, \\
\bb^{(2)}\!&:=q_1 \overline{x}_1^2 + l_1\overline{x}_1  + c_1 = \varphi_1(\overline{x}_1),
\end{align*}
and the augmented input $\xib(x):= \begin{pmatrix}
x & x^2 \end{pmatrix}^\top$, which is such that
$\varphi(x)=\Phib(\xib(x))$ for every $x \in [\underline{x}_1,\overline{x}_s]$, where
$$
\kappa_i:=l_i + q_i (\underline{x}_i+\overline{x}_i)
$$
for every $i \in \{1,\dots,s\}$. 

\section{Tailored max-out-NN for PWQ functions}\label{sec:tailoredNN}

While the approach summarized in Section~\ref{subsec:tailoredReLU} works well for finding tailored ReLU-NN for PWQ functions with $n=1$, the example in \cite[Sect.~IV-D]{Teichrib2021} shows that an extension to higher dimensional problems (with $n>1$) is not straightforward. In fact, the central proof of \cite[Thm.~1]{Teichrib2021}
makes use of ReLU-neurons that are ``active'' only for $\xb$ inside a certain region $\Rc^{(i)}$. While such a construction can easily be derived for $n=1$,  an extension to $n>1$ is hard if not impossible. This observation motivates the work at hand, where we aim for an exact representation of convex PWQ functions that applies for $n\geq 1$. However, similar to the approach in \cite{Teichrib2021}, we will initially focus on the special case $n=1$. Nevertheless, we will illustrate that an extension to $n>1$ is within reach with a numerical example in Section~\ref{subsec:ex_2D}.

The central idea here is to exploit the convexity of the PWQ function more explicitly than in \cite[Thm.~1]{Teichrib2021}. In this context, it is well-known that a convex PWA function $h(\xb)$ can be evaluated by computing the maximum of all affine segments $h_i(\xb)$ (independent of their various domains), i.e.,
\begin{equation}
\label{eq:maxPWA}
h(\xb)=\max\limits_{1\leq i \leq s}\{h_i(\xb)\}.
\end{equation}
It is easy to see that this operation is equivalent to the evaluation of a max-out neuron. Unfortunately, the relation
\begin{equation}
\label{eq:maxQuad}
\varphi(\xb)=\max\limits_{1\leq i \leq s}\{\varphi_i(\xb)\},
\end{equation}
which is inspired by \eqref{eq:maxPWA}, does not hold in general. In fact, Figure~\ref{fig:phi} shows a convex PWQ function for which~\eqref{eq:maxQuad} is violated since some quadratic segments ``dominate'' outside their domains. The leading idea now is to compensate this defect by adding a suitable convex PWA function $h(\xb)$ to $\varphi(\xb)$ such that
\begin{equation}
\label{eq:maxQuadPlusH}
\varphi(\xb)+h(\xb)=\max\limits_{1\leq i \leq s}\{\varphi_i(\xb)+h_i(\xb)\}
\end{equation}
applies for all $\xb$ in the domain of $\varphi$. Clearly, this would immediately allow to represent $\varphi$ in terms of a max-out network with one hidden layer and two neurons. In fact, this follows from the trivial observation that 
$$
\varphi(\xb)=\underbrace{\varphi(\xb)+h(\xb)}_{\text{first neuron}}\,\,-\!\!\!\!\underbrace{h(\xb)}_{\text{second neuron}}\!\!\!\!,
$$
where the two neurons reflect the right-hand sides in \eqref{eq:maxPWA} and~\eqref{eq:maxQuadPlusH}.
Based on the previous discussion, we formulate the following conjecture that will guide us through the remaining paper.

\begin{conj}
\label{conj:maxoutPWQ}
Every convex PWQ function can be exactly represented by a max-out-NN with one hidden layer and two neurons.
\end{conj}

Before deriving a proof of the conjecture for the special case $n=1$, we briefly note that it can be considered as a tailored extension of the central observation in \cite{Goodfellow2013}. In fact, \cite[Thm.~4.3]{Goodfellow2013} states that that a max-out-NN with one hidden layer and two neurons allows to approximate any continous function arbitrarily well (but not necessarily exactly).

Now, specifying the ideas from above for $n=1$ leads to the consideration of convex PWQ functions of the form \eqref{eq:PWQ1D} and convex PWA functions
\begin{equation}\label{eq:h}
h(x) := \left\{
\begin{array}{cc}
\alpha_1 x +\beta_1 & \text{if}\,\,\,x\in[\underline{x}_1,\overline{x}_1], \\
\vdots &  \vdots \\
\alpha_s x +\beta_s & \text{if}\,\,\,x\in[\underline{x}_s,\overline{x}_s]. \\
\end{array}
\right.\!
\end{equation}
Analogously to~\eqref{eq:convAndContPhi}, continuity and convexity of $h$ can now easily be specified by the conditions
\begin{subequations}
\label{eq:PWAandConvex}
\begin{align}
    \alpha_i\overline{x}_i +\beta_i &= \alpha_{i+1}\overline{x}_i +\beta_{i+1} \quad \text{and}  \\
    \alpha_i &\leq \alpha_{i+1},  
\end{align}
\end{subequations}
for every $i\in\{1,\dots,s-1\}$, respectively.
Now, aiming for the relation \eqref{eq:maxQuadPlusH}, we pose the additional constraints
\begin{subequations}
\label{eq:conditionsLeftOfi}
\begin{align}
\label{eq:jBelowiAtLowerIntervalBound}
  \varphi_j(\underline{x}_i) +\alpha_j\underline{x}_i +\beta_j &\leq   \varphi_i(\underline{x}_i)+ \alpha_i\underline{x}_i +\beta_i  \\
  \nonumber
    \varphi_j(\overline{x}_i) +\alpha_j\overline{x}_i +\beta_j &\leq   \varphi_i(\underline{x}_i)+ \alpha_i\underline{x}_i +\beta_i \\
    \label{eq:jBelowTangentAtUpperIntervalBound}
    &\quad+(\overline{x}_i-\underline{x}_i)(2 q_i \underline{x}_i + l_i +\alpha_i)
\end{align}
\end{subequations}
for every $i\in\{2,\dots,s\}$ and $j \in \{1,\dots,i-1\}$ as well as
\begin{subequations}
\label{eq:conditionsRightOfi}
\begin{align}
\label{eq:jBelowiAtUpperIntervalBound}
  \varphi_j(\overline{x}_i) +\alpha_j\overline{x}_i +\beta_j  &\leq   \varphi_i(\overline{x}_i)+ \alpha_i\overline{x}_i +\beta_i  \\
  \nonumber
    \varphi_j(\underline{x}_i) +\alpha_j\underline{x}_i +\beta_j &\leq   \varphi_i(\overline{x}_i)+ \alpha_i\overline{x}_i +\beta_i \\
    \label{eq:jBelowTangentAtLowerIntervalBound}
    &\quad-(\overline{x}_i-\underline{x}_i)(2 q_i \overline{x}_i + l_i +\alpha_i)
\end{align}
\end{subequations}
for every $i\in\{1,\dots,s-1\}$ and $j \in \{i+1,\dots,s\}$. The underlying concepts will be clarified in the proof of the following theorem.

\begin{figure}
\centering
    \subfigure[\hspace{-6mm} \label{fig:conditionsLeftOfi}]{\includegraphics[trim={6.5cm 11cm 6cm 11cm},clip,scale=0.5]{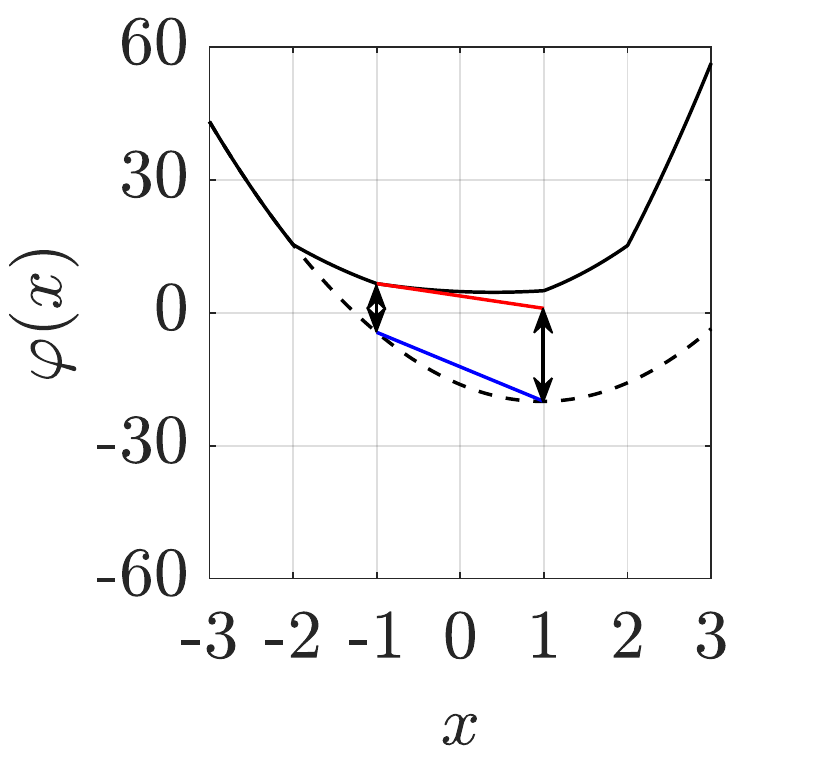}}
    \subfigure[\hspace{-6mm} \label{fig:conditionsRightOfi}]{\includegraphics[trim={6.5cm 11cm 6cm 11cm},clip,scale=0.5]{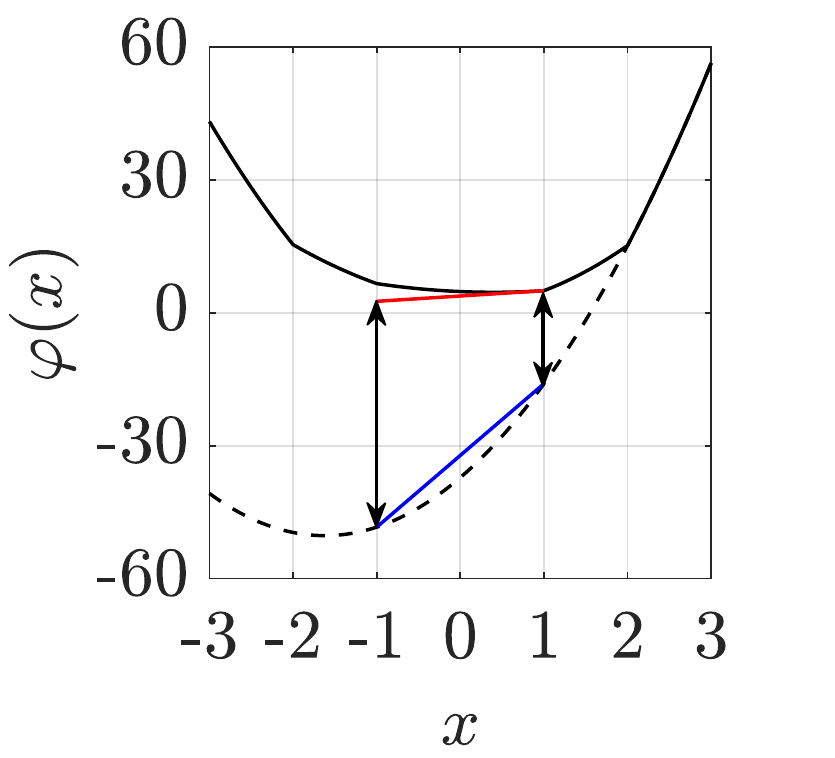}}
\caption{Exemplary illustration of the conditions \eqref{eq:conditionsLeftOfi} and \eqref{eq:conditionsRightOfi} for $i=3$ and $j\in\{1,5\}$. In (a), \eqref{eq:conditionsLeftOfi} is shown for $j=1<i$ with the tangent on $\varphi(\underline{x}_i)+h(\underline{x}_i)$ in red evaluated at the points $\underline{x}_i$ and $\overline{x}_i$.
In (b), \eqref{eq:conditionsRightOfi} is illustrated for $j=5>i$,
where the red line is the tangent on $\varphi(\overline{x}_i)+h(\overline{x}_i)$ evaluated at the point $\underline{x}_i$ and $\overline{x}_i$. In both plots, the blue line illustrates the interpolation of $\varphi_j(\hat{x})+\alpha_j \hat{x} +\beta_j$ between the points $\underline{x}_i$ and $\overline{x}_i$.}\label{fig:conditions}

\end{figure}

\begin{thm}\label{thm:sumEqualsMaxOutNeuron}
Assume there exist $\alpha_1,\dots,\alpha_s \in \R$ and $\beta_1,\dots,\beta_s \in \R$ satisfying the constraints~\eqref{eq:PWAandConvex}--\eqref{eq:conditionsRightOfi}. Then,
\begin{equation}
\label{eq:sumEqualsMaxOutNeuron}
 \varphi(x)+h(x)=\max_{1\leq i \leq s} \{ \varphi_i(x)+\alpha_i x +\beta_i  \}.
\end{equation}
%holds.
\end{thm}

\vspace{2mm}
\begin{proof}
To prove~\eqref{eq:sumEqualsMaxOutNeuron}, we consider any $i\in\{1,\dots,s\}$ and $\hat{x} \in [\underline{x}_i,\overline{x}_i]$. By definition of $\varphi$ and $h$, we then obtain
$$
\varphi(\hat{x})+h(\hat{x}) = \varphi_i(\hat{x})+\alpha_i \hat{x} +\beta_i.
$$
Since the segment $i$ also appears on the right-hand side of~\eqref{eq:sumEqualsMaxOutNeuron}, we find
$$
\varphi(\hat{x})+h(\hat{x}) \leq \max_{1\leq i \leq s} \{ \varphi_i(\hat{x})+\alpha_i \hat{x} +\beta_i  \}
$$
by construction. Hence, \eqref{eq:sumEqualsMaxOutNeuron} can only be violated if there exists a $j\in\{1,\dots,s\} \setminus \{i\}$ such that 
\begin{equation}
\label{eq:toBeContradicted}
\varphi_i(\hat{x})+\alpha_i \hat{x} +\beta_i < \varphi_j(\hat{x})+\alpha_j \hat{x} +\beta_j,
\end{equation}
which, however, leads to a contradiction. To see this, we distinguish the two cases (I) $j<i$ and (II) $i<j$. 
Regarding the first case, we initially note that \eqref{eq:toBeContradicted} would require $\hat{x}>\underline{x}_i$ to comply with~\eqref{eq:jBelowiAtLowerIntervalBound}. Hence,
 $$
 \eta:=\frac{\hat{x}-\underline{x}_i}{\overline{x}_i-\underline{x}_i} \in (0,1].
 $$
Further, due to convexity of segments $i$ and $j$, we require
 \begin{subequations}
 \begin{align}
 \nonumber
  \varphi_i(\hat{x})+\alpha_i \hat{x} +\beta_i &\geq   \varphi_i(\underline{x}_i)+ \alpha_i\underline{x}_i +\beta_i \\
  \label{eq:iHatAboveTangent}
   &\quad+(\hat{x}-\underline{x}_i)(2 q_i \underline{x}_i + l_i +\alpha_i), \\
   \nonumber
   \varphi_j(\hat{x})+\alpha_j \hat{x} +\beta_j & \leq (1-\eta)\left(\varphi_j(\underline{x}_i) +\alpha_j\underline{x}_i +\beta_j\right) \\
   \label{eq:jHatBelowInterpolation}
    &\quad+ \eta\left(\varphi_j(\overline{x}_i) +\alpha_j\overline{x}_i +\beta_j\right).
 \end{align}
 \end{subequations}
 In fact, $\varphi_i(\hat{x})+\alpha_i \hat{x} +\beta_i$ is restricted to lie above (or at) any tangent of segment $i$ and the right-hand side of~\eqref{eq:iHatAboveTangent} reflects the tangent at $\underline{x}_i$ (see the red line in Figure~\ref{fig:conditions}). Analogously $\varphi_j(\hat{x})+\alpha_j \hat{x} +\beta_j$ has to lie below (or at) any interpolation between two points on segment $j$ and the right-hand side of~\eqref{eq:jHatBelowInterpolation} reflects the interpolation between the supporting points $\underline{x}_i$ and $\overline{x}_i$ (which is illustrated by the blue line in Figure~\ref{fig:conditions}). Now, multiplying \eqref{eq:jBelowTangentAtUpperIntervalBound} with $\eta$
 results in
\begin{align*}
    \eta \left(\varphi_j(\overline{x}_i) +\alpha_j\overline{x}_i +\beta_j\right) &\leq   \eta \left( \varphi_i(\underline{x}_i)+ \alpha_i\underline{x}_i +\beta_i \right) \\
    &\quad+(\hat{x}-\underline{x}_i)(2 q_i \underline{x}_i + l_i +\alpha_i)
\end{align*}
Next, we use~\eqref{eq:iHatAboveTangent} to overestimate the right-hand side of the former relation and substitute the result on the right-hand side of~\eqref{eq:jHatBelowInterpolation} in order to obtain
 \begin{align*}
   \varphi_j(\hat{x})+\alpha_j \hat{x} +\beta_j & \leq (1-\eta)\left(\varphi_j(\underline{x}_i) +\alpha_j\underline{x}_i +\beta_j\right) \\
    &\quad- (1-\eta)\left(\varphi_i(\underline{x}_i)+\alpha_i\underline{x}_i +\beta_i\right) \\
    &\quad+\varphi_i(\hat{x})+\alpha_i \hat{x} +\beta_i.
 \end{align*}
Finally, taking~\eqref{eq:jBelowiAtLowerIntervalBound} and $1-\eta \in [0,1)$ into account, we find
$$
\varphi_j(\hat{x})+\alpha_j \hat{x} +\beta_j \leq \varphi_i(\hat{x})+\alpha_i \hat{x} +\beta_i,
$$
which indeed contradicts~\eqref{eq:toBeContradicted}. Since the second case can be handled analogously, the proof is complete.
\end{proof}

\begin{rem}\label{rem:OP}
Since the conditions~\eqref{eq:PWAandConvex}--\eqref{eq:conditionsRightOfi} required to guarantee \eqref{eq:sumEqualsMaxOutNeuron} are affine in $\alpha_i$ and $\beta_i$, we might be able to compute a suitable PWA function $h(x)$ by solving an optimization problem (OP) of the form
\begin{align}\label{eq:OP}
\min_{\substack{\alpha_1,\beta_1,\dots,\alpha_s,\beta_s}} &J\left(\alpha_1,\beta_1,\dots,\alpha_s,\beta_s\right)   \\
\nonumber
\text{s.t.} \qquad & \text{\eqref{eq:PWAandConvex}--\eqref{eq:conditionsRightOfi}\,\,\,(for $i$ and $j$ as above).}
\end{align}
Here, the cost function $J$ can be chosen to enforce certain shapes of $h$. For instance, if the impact of $h$ should be as small as possible, $\sum_{i=1}^s \alpha_i^2+\beta_i^2$ is a suitable choice for $J$, which results in a quadratic program (QP).
\end{rem}
It remains to comment on the feasibility of the constraints \eqref{eq:PWAandConvex}--\eqref{eq:conditionsRightOfi} and, hence, the feasibility of \eqref{eq:OP}.
In this context, Theorem~\ref{thm:feasibleSolution} further below states the constraints are always feasible since a feasible solution can always be computed according to Algorithm~\ref{alg:feasibleSolution}.

\begin{alg} Feasible solution to \eqref{eq:PWAandConvex}--\eqref{eq:conditionsRightOfi}.
\label{alg:feasibleSolution}

\begin{algorithmic}[1]
\State initialize $\alpha_i \gets 0$ and $\beta_i \gets 0$ for every $i \in \{1,\dots,s\}$
\For{$i=1,\dots,s$}
\State  set the auxiliary quantities $\gamma_1 \gets \varphi_i(\underline{x}_i)+ \alpha_i\underline{x}_i +\beta_i$, 
\Statex \quad\,\,\,$\gamma_2\gets \gamma_1+(\overline{x}_i-\underline{x}_i) (2 q_i \underline{x}_i+l_i+\alpha_i)$, $\gamma_3\gets \varphi_i(\overline{x}_i)+$
\Statex  \quad\,\,\,$\alpha_i\overline{x}_i +\beta_i$, and $\gamma_4 \gets \gamma_3 - (\overline{x}_i-\underline{x}_i) (2 q_i \overline{x}_i+l_i+\alpha_i)$
\For{$j=1,\dots,i-1$}
\State set $\Delta \alpha \gets 0$
\If{\eqref{eq:jBelowiAtLowerIntervalBound} is violated} %\Comment{requires $j<i-1$}
\State Set $\Delta \alpha \gets  \dfrac{\gamma_1  - \varphi_j(\underline{x}_i) -\alpha_j\underline{x}_i -\beta_j }{\underline{x}_i-\overline{x}_j}$
\EndIf
\If{\eqref{eq:jBelowTangentAtUpperIntervalBound} is violated} 
\State $\Delta \alpha \leftarrow \min\left\{\!\Delta \alpha , \dfrac{\gamma_2 -\varphi_j(\overline{x}_i) -\alpha_j\overline{x}_i -\beta_j}{\overline{x}_i-\overline{x}_j} \! \right\}$
\EndIf
\If{$\Delta \alpha<0$} 
\State perform the updates $\alpha_k \gets \alpha_k+\Delta \alpha$ and 
\Statex \qquad\qquad\, $\beta_k \gets \beta_k-\Delta \alpha \overline{x}_j$ for every $k\in \{1,\dots,j\}$
\EndIf
\EndFor
\For{$j=i+1,\dots,s$}
\State set $\Delta \alpha \gets 0$
\If{\eqref{eq:jBelowiAtUpperIntervalBound} is violated} %\Comment{requires $j>i+1$}
\State Set $\Delta \alpha \gets  \dfrac{\varphi_j(\overline{x}_i) +\alpha_j\overline{x}_i +\beta_j -  \gamma_3 }{\underline{x}_j-\overline{x}_i}$
\EndIf
\If{\eqref{eq:jBelowTangentAtLowerIntervalBound} is violated} 
\State $\Delta \alpha \leftarrow \max\left\{ \!\Delta \alpha , \dfrac{\varphi_j(\underline{x}_i) +\alpha_j\underline{x}_i +\beta_j-\gamma_4}{\underline{x}_j-\underline{x}_i} \! \right\}$
\EndIf
\If{$\Delta \alpha>0$} 
\State perform the updates $\alpha_k \gets \alpha_k+\Delta \alpha$ and 
\Statex \qquad\qquad\, $\beta_k \gets \beta_k-\Delta \alpha \underline{x}_j$ for every $k\in \{j,\dots,s\}$
\EndIf
\EndFor
\EndFor
\State \Return $\alpha_1,\dots,\alpha_s$ and $\beta_1,\dots,\beta_s$
\end{algorithmic}
\end{alg}

\begin{thm}\label{thm:feasibleSolution}
Algorithm~\ref{alg:feasibleSolution} provides a feasible solution to the constraints \eqref{eq:PWAandConvex}--\eqref{eq:conditionsRightOfi}. 
\end{thm}

\begin{proof}
For every $i$ and $j\neq i$, we have to satisfy a pair of constraints in \eqref{eq:conditionsLeftOfi} and \eqref{eq:conditionsRightOfi}. Obviously, Alg. 1 runs through these constraints iteratively. Let us take a snapshot of the algorithms for some $i$ and $j\neq i$. Clearly, after evaluating the corresponding body of the for-loops, the corresponding pair of constraints holds. If we can prove that all previously addressed constraints still hold as well, we are done.

Assume the constraints have already been satisfied for the pair $(\hat{\imath},\hat{\jmath})$ and show that the modifications associated with $(i,j)$ will not alter this. Since we have $\hat{\imath} \neq\hat{\jmath}$ and $i\neq j$, we can distinguish the four cases (1) $\hat{\imath}<\hat{\jmath}$ and $i<j$, (2) $\hat{\imath}<\hat{\jmath}$ and $j<i$, (3) $\hat{\jmath}<\hat{\imath}$ and $i<j$, and (4) $\hat{\jmath}<\hat{\imath}$ and $j<i$. 
We next prove the second case as it nicely illustrates the involved steps.
Since this case offers $\hat{\imath}<\hat{\jmath}$, the conditions
\begin{subequations}
\label{eq:satisfiedConditions}
\begin{align}
  \varphi_{\hat{\jmath}}(\overline{x}_{\hat{\imath}}) +\alpha_{\hat{\jmath}}\overline{x}_{\hat{\imath}} +\beta_{\hat{\jmath}}  &\leq   \varphi_{\hat{\imath}}(\overline{x}_{\hat{\imath}})+ \alpha_{\hat{\imath}}\overline{x}_{\hat{\imath}} +\beta_{\hat{\imath}}  \\
      \nonumber
    \varphi_{\hat{\jmath}}(\underline{x}_{\hat{\imath}}) +\alpha_{\hat{\jmath}}\underline{x}_{\hat{\imath}} +\beta_{\hat{\jmath}} &\leq   \varphi_{\hat{\imath}}(\overline{x}_{\hat{\imath}})+ \alpha_{\hat{\imath}}\overline{x}_{\hat{\imath}} +\beta_{\hat{\imath}}\\
    &\quad-(\overline{x}_{\hat{\imath}}-\underline{x}_{\hat{\imath}})(2 q_{\hat{\imath}} \overline{x}_{\hat{\imath}} + l_{\hat{\imath}} +\alpha_{\hat{\imath}})
\end{align}
\end{subequations}
hold before the modifications. Due to $j<i$, the modifications will affect the parameters $\alpha_1,\dots,\alpha_j$ and $\beta_1,\dots,\beta_j$. Clearly, the modified indices may or may not involve $\hat{\imath}$ and $\hat{\jmath}$. Hence, it is reasonable to distinguish the subcases (2.1) $j<\hat{\imath}$, (2.2) $\hat{\imath}\leq j < \hat{\jmath}$, and (2.3) $ \hat{\jmath}\leq j$. Taking $\hat{\imath}<\hat{\jmath}$ into account, the first subcase implies that modifications do not affect the conditions \eqref{eq:satisfiedConditions}. In the second subcase, we obtain
\begin{align*}
  \varphi_{\hat{\jmath}}(\overline{x}_{\hat{\imath}}) +\alpha_{\hat{\jmath}}\overline{x}_{\hat{\imath}} +\beta_{\hat{\jmath}}  &\leq   \varphi_{\hat{\imath}}(\overline{x}_{\hat{\imath}})+ (\alpha_{\hat{\imath}}+\Delta \alpha) \overline{x}_{\hat{\imath}} +\beta_{\hat{\imath}} - \Delta \alpha \overline{x}_j \\
    \varphi_{\hat{\jmath}}(\underline{x}_{\hat{\imath}}) +\alpha_{\hat{\jmath}}\underline{x}_{\hat{\imath}} +\beta_{\hat{\jmath}} &\leq   \varphi_{\hat{\imath}}(\overline{x}_{\hat{\imath}})+ (\alpha_{\hat{\imath}}+\Delta \alpha)\overline{x}_{\hat{\imath}} +\beta_{\hat{\imath}} - \Delta \alpha \overline{x}_j \\
    &\quad-(\overline{x}_{\hat{\imath}}-\underline{x}_{\hat{\imath}})(2 q_{\hat{\imath}} \overline{x}_{\hat{\imath}} + l_{\hat{\imath}} +\alpha_{\hat{\imath}}+\Delta \alpha).
\end{align*}
The conditions remain valid since the modifications only increase the right-hand sides of~\eqref{eq:satisfiedConditions}. In fact, the novel terms 
\begin{subequations}
\begin{align}
\label{eq:firstNovelTerm}
\Delta \alpha \overline{x}_{\hat{\imath}}-\Delta \alpha \overline{x}_j &= -\Delta \alpha (  \overline{x}_j-  \overline{x}_{\hat{\imath}}) \quad \text{and} \\
\label{eq:secondNovelTerm}
-\Delta \alpha(\overline{x}_{\hat{\imath}}-\underline{x}_{\hat{\imath}}) &
\end{align}
\end{subequations}
are non-negative due to the relations $\Delta \alpha<0$, $\overline{x}_j-  \overline{x}_{\hat{\imath}}\geq 0$, and $\overline{x}_{\hat{\imath}}-\underline{x}_{\hat{\imath}}>0$. Now, in the third subcase, the term~\eqref{eq:firstNovelTerm} also appears on the left-hand sides and, hence, the corresponding terms cancel out. The remaining term~\eqref{eq:secondNovelTerm} is non-negative as before and, consequently, the conditions remain valid.

We will leave the remaining cases for the interested reader. However, we briefly note that, taking the order of iterations in Algorithm~\ref{alg:feasibleSolution} into account, we find the additional condition $\hat{\imath}\leq i$. As a consequence, case (3) can be specified to $\hat{\jmath}<\hat{\imath} \leq i<j$. Under this specification, it quickly turns out that any modifications associated with this case does not alter the conditions obtained for $(\hat{\imath},\hat{\jmath})$. Hence, only the first and fourth case remain open (but can be easily handled analogously to the second case).
\end{proof}

The combination of Theorems~\ref{thm:sumEqualsMaxOutNeuron} and \ref{thm:feasibleSolution} leads to the following summarizing statement about the structure of a tailored max-out-NN for \eqref{eq:PWQ1D} that is in line with Conjecture~\ref{conj:maxoutPWQ}.

\begin{cor}\label{cor:ANNforPhi}
Every convex PWQ of the form~\eqref{eq:PWQ1D} %with scalar variables 
can be exactly represented by a max-out-NN with one hidden layer of width $w_1=2$, $p_1=s$, and %the augmented input 
$\xib:=(
x \,\,\,\, x^2 )^\top$.
\end{cor}

\begin{proof}
A max-out-NN with the specified structure is given by
\begin{align*}
\Phib(\xib)=\Wb^{(2)}\begin{pmatrix}
\max\limits_{1\leq i \leq s}\big\{\Wb^{(1)}_{i}\xib+\bb^{(1)}_{i}\big\} \\
\max\limits_{s+1\leq i \leq 2s}\big\{\Wb^{(1)}_{i}\xib+\bb^{(1)}_{i}\big\}
\end{pmatrix}+\bb^{(2)},
\end{align*}
where $\Wb^{(1)}_i$ denotes the $i$-th row of $\Wb^{(1)}$.
Now, we assume that $\alpha_i$ and $\beta_i$ have been chosen such \eqref{eq:PWAandConvex}-\eqref{eq:conditionsRightOfi} hold, which is always possible according to Theorem~\ref{thm:feasibleSolution}.
Then, specifying the weights and biases as
\begin{subequations}\label{eq:WbforPhi}
\begin{align}
    \Wb^{(1)}&:=\begin{pmatrix} 
\alpha_1+l_1 & q_1 \\ 
\vdots & \vdots \\ 
\alpha_s+l_s & q_s  \\ 
\alpha_1 & 0\\ 
\vdots & \vdots \\
\alpha_1 & 0 \\ 
\end{pmatrix}, &  \bb^{(1)}&:=\begin{pmatrix}
\beta_1+c_1 \\
\vdots \\
\beta_s+c_s  \\
\beta_1 \\
\vdots \\
\beta_s
\end{pmatrix}, \\
\Wb^{(2)}&:=\begin{pmatrix} 
1 & -1
\end{pmatrix} & \bb^{(2)}&:=0
\end{align}
\end{subequations}
results in
\begin{equation}
\label{eq:PhiEvaluated}
    \Phib(\xib)=\max\limits_{1 \leq i \leq s}\{\varphi_i(x)  + \alpha_i x + \beta_i\}-\max\limits_{1\leq i \leq s}\{\alpha_i x + \beta_i\}.
\end{equation}
This is equivalent to $\varphi(x)$ since the right-hand side in~\eqref{eq:PhiEvaluated} evaluates to $(\varphi(x)+h(x))-h(x)=\varphi(x)$ according to Theorem~\ref{thm:sumEqualsMaxOutNeuron} and due to convexity of $h(x)$.
\end{proof}

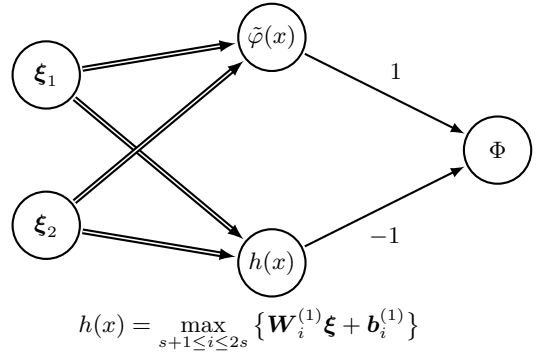
\begin{figure}[ht]
\centering
\begin{tikzpicture}[thick,scale=1, every node/.style={scale=1}]
	\small

 \node [draw,circle,scale=3]   (I1) at (0,1) {};
 
 \node [draw,circle,scale=3]   (I2) at (0,-1){}; 
 
 \node [draw,circle,scale=3]   (L1) at (3,1.5) {};
 
 \node [draw,circle,scale=3]   (L2) at (3,-1.5){};  
 
 \node [draw,circle,scale=3]   (O1) at (6,0){}; 
 
  \node at (0,1) {$\xib_1$};
  \node at (0,-1) {$\xib_2$};
   \node at (2.7,2.4) {$\tilde{\varphi}(x):=\varphi(x)+h(x)=\max\limits_{1\leq i \leq s}\big\{\Wb^{(1)}_{i}\xib+\bb^{(1)}_{i}\big\}$};
   \node at (2.7,-2.4) {$h(x)=\max\limits_{s+1\leq i \leq 2s}\big\{\Wb^{(1)}_{i}\xib+\bb^{(1)}_{i}\big\}$};
  
   \node at (4.65,1.05) {$1$};
   \node at (4.5,-1.15) {$-1$};

\node at (3,1.5) {$\tilde{\varphi}(x)$};
\node at (3,-1.5) {$h(x)$};
  
\node at (6,0) {$\Phi$};

	\draw [thick, double, ->, >=latex] (I1) to (L1);
	\draw [thick, double, ->, >=latex] (I1) to (L2);
	\draw [thick, double, ->, >=latex] (I2) to (L1);
	\draw [thick, double, ->, >=latex] (I2) to (L2);
	
	\draw [thick, ->, >=latex] (L1) to (O1);	
	\draw [thick, ->, >=latex] (L2) to (O1);
	
	\end{tikzpicture}
\caption{Illustration of the max-out-NN according to Corollary~\ref{cor:ANNforPhi} for an exact representation of the convex PWQ function $\varphi$. Double arrows highlight multi-channel preactivations with $p_1=s$.}
\label{fig:maxoutTwoNeurons}
\end{figure}

Figure~\ref{fig:maxoutTwoNeurons} shows the structure of the resulting max-out-NN with one neuron for the computation of the PWQ function $\varphi(x)+h(x)$ and one neuron for the computation of the PWA function $h(x)$.

\section{Numerical examples}
\label{sec:example}
We illustrate our results with two numerical examples. The first examples demonstrate the application of Theorem~\ref{thm:sumEqualsMaxOutNeuron} and Corollary~\ref{cor:ANNforPhi}. The  second example shows that the methodology can, in principle, also be applied for $n>1$.

\subsection{1D PWQ function}\label{subsec:ex_1D}
\begin{figure}[h]
\centering
\includegraphics[trim={2cm 11cm 3cm 11cm},clip,scale=0.5]{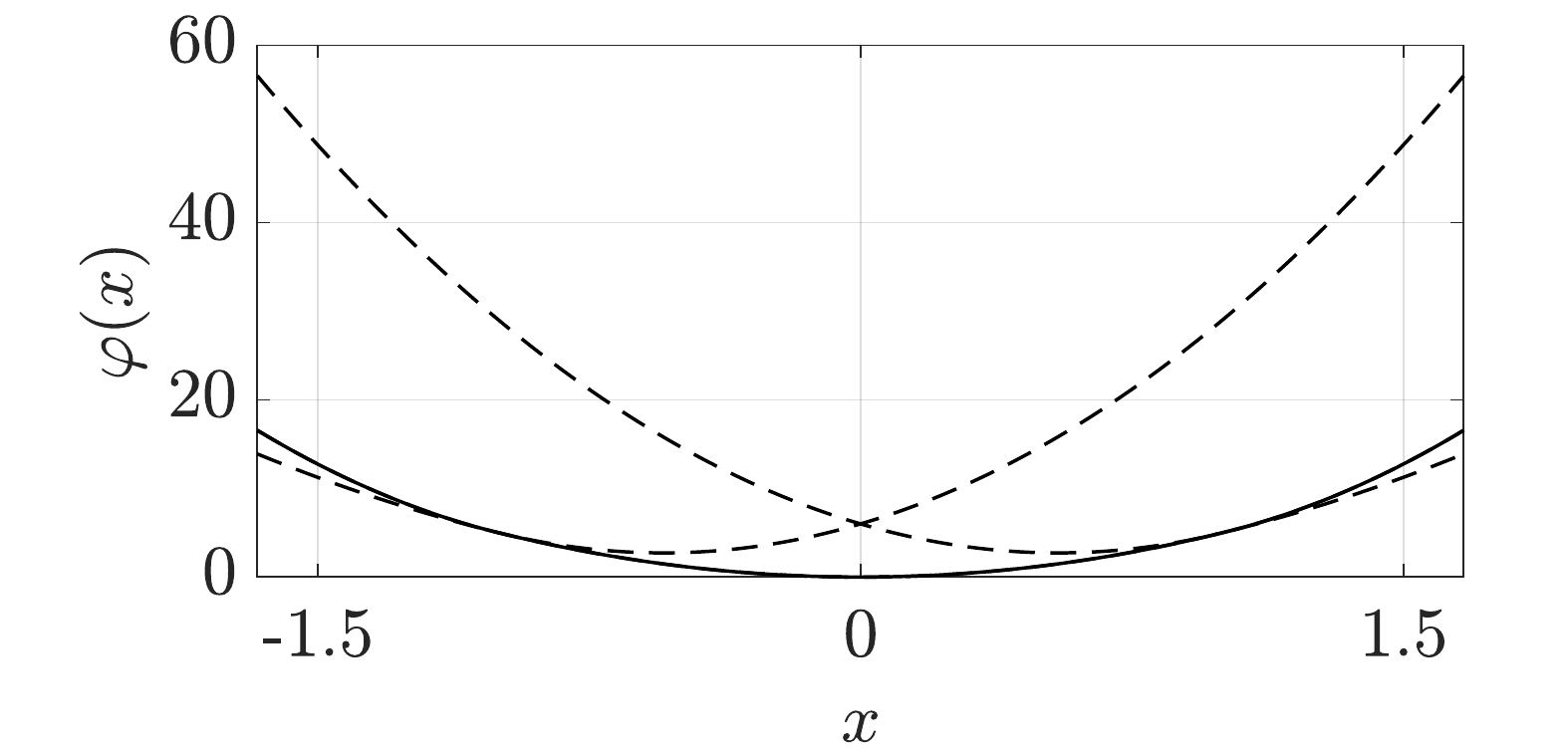}
\caption{Illustration of the PWQ function $\varphi(x)$ in black and the extension of the quadratic segments outside their regions in black dashed.}
\label{fig:phi}
\end{figure}

Explicitly solving the optimal control problem (OCP)
\begin{align*}
\varphi(x) :&= \min_{\substack{x_0,x^+,u}} 5(x^+)^2 + \tfrac{19}{5} x_0^2 + u^2 \\
\text{s.t.} \quad x_0&=x \\
x^+&=\tfrac{6}{5}x_0+u \\
\quad  x_0 &\in [-10,10], \ x^+  \in [-1,1], \ u  \in [-1,1] ,
\end{align*}
where $x$ is the current system state results in the OVF
\begin{equation}\label{eq:example1_V1}
\varphi(x) = \left\{
\begin{array}{lll}
11x^2+12x+6 & \text{if} & x\in\left[-\tfrac{5}{3},-1\right], \\
5x^2 & \text{if} & x\in\left[-1,1\right], \\
11x^2-12x+6 & \text{if} & x\in\left[1,\tfrac{5}{3}\right], 
\end{array}
\right. 
\end{equation}
where we refer to~\cite[Ex.~2]{SchulzeDarup2016_ECC_MPC} for details on the underlying MPC problem and system.
Now, Figure~\ref{fig:phi} shows that we have
\begin{align*}
\varphi(x) \neq \max\limits_{1\leq i\leq 3}\left\{\varphi_i(x)\right\},
\end{align*}
since the segment $\varphi_2(x)$ is smaller than $\varphi_1(x)$ and $\varphi_3(x)$ for all $x \in (-1,1)$. Hence, $\varphi$ does not offer a trivial representation by one max-out neuron.
\vspace{-2mm}

\begin{figure}[h]
\centering
    \subfigure[\hspace{-6mm} \label{fig:phi+h_alg}]{\includegraphics[trim={6.5cm 10.9cm 6cm 11cm},clip,scale=0.5]{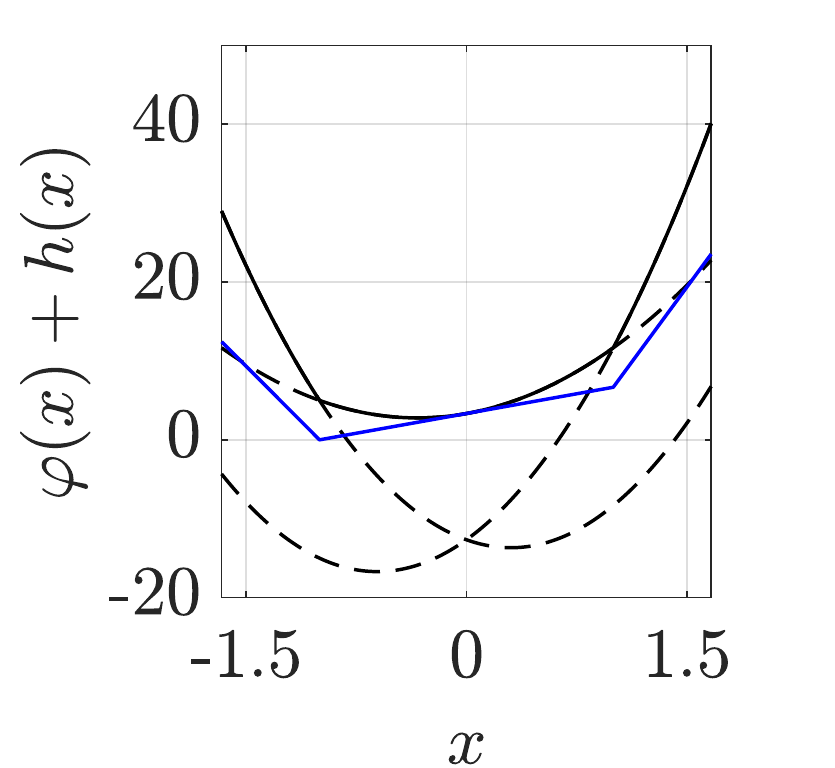}}
    \subfigure[\hspace{-6mm} \label{fig:phi+h_OP}]{\includegraphics[trim={6.5cm 10.9cm 6cm 11cm},clip,scale=0.5]{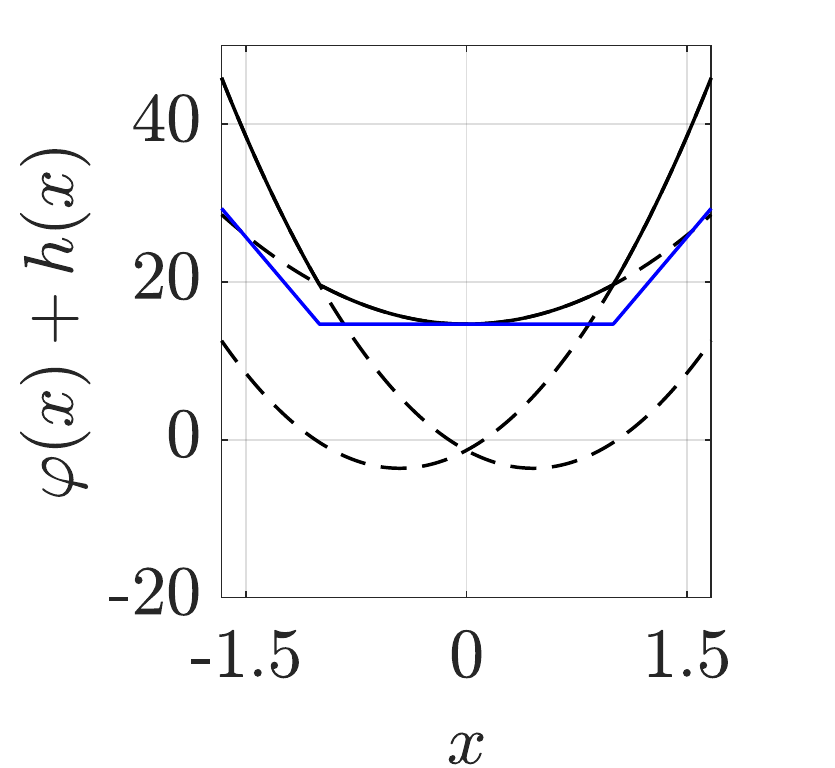}}
\caption{Illustration of the PWQ function $\varphi(x)+h(x)$ in black, the extension of the quadratic segments outside their regions in black dashed, and the PWA function $h(x)$ in blue. The function $h(x)$ is illustrated for $\alpha_i=\tilde{\alpha}_i$, $\beta_i=\tilde{\beta}_i$ in (a) and for $\alpha_i=\hat{\alpha}_i$, $\beta_i=\hat{\beta}_i$ in (b) with $1\leq i \leq 3$.}
\label{fig:phi+h}
\end{figure}

However, an exact representation in terms of a max-out-NN can be computed according to the proposed procedure in Section~\ref{sec:tailoredNN}. To this end, we initially apply Algorithm~\ref{alg:feasibleSolution} in order to compute the parameters
\begin{subequations}\label{eq:alphabetaAlg}
\begin{align}
\tilde{\alpha}_1&=-\frac{56}{3},\quad\tilde{\alpha}_2=\frac{10}{3},\quad\tilde{\alpha}_3=\frac{76}{3},\\
\tilde{\beta}_1&=-\frac{56}{3},\quad\tilde{\beta}_2=\frac{10}{3},\quad\tilde{\beta}_3=-\frac{56}{3}
\end{align}
\end{subequations}
that satisfy the conditions \eqref{eq:PWAandConvex}--\eqref{eq:conditionsRightOfi}. Thus we have
\begin{align}
\nonumber
\varphi(x)=\max\big\{&11x^2-\tfrac{20}{3}x-\tfrac{38}{3},5x^2+\tfrac{10}{3}x+\tfrac{10}{3},\\
\nonumber
&11x^2+\tfrac{40}{3}x-\tfrac{38}{3}\big\}\\
\label{eq:examplePhiMax}
-\max\big\{&-\tfrac{56}{3}x - \tfrac{56}{3},\tfrac{10}{3}x + \tfrac{10}{3}, \tfrac{76}{3}x - \tfrac{56}{3}\big\}.
\end{align}
Figure~\ref{fig:phi+h}(a) shows that in contrast to the function $\varphi(x)$ (Figure~\ref{fig:phi}), the quadratic segments of the function $\varphi(x)+h(x)$ are such that they are greater than all other segments inside their region which allows the representation of \eqref{eq:examplePhiMax} in the form \eqref{eq:sumEqualsMaxOutNeuron} for $\alpha_i=\tilde{\alpha}_i$ and $\beta_i=\tilde{\beta}$.

As mentioned in Remark~\ref{rem:OP} the function $h$ can also be computed as a solution of the OP \eqref{eq:OP}. Therefore we consider in this example the cost function 
\begin{equation}\label{eq:cost_example}
J\left(\hat{\alpha}_1,\hat{\beta}_1,\dots,\hat{\alpha}_s,\hat{\beta}_s\right)=\sum\limits_{i=1}^s\hat{\alpha}_i^2+\hat{\beta}_i^2.
\end{equation}
The parameters
\begin{subequations}\label{eq:alphabetaOP}
\begin{align}
\hat{\alpha}_1&=-22,\quad\hat{\alpha}_2=0,\quad\hat{\alpha}_3=22,\\
\hat{\beta}_1&=-\frac{22}{3},\quad\hat{\beta}_2=\frac{44}{3},\quad\hat{\beta}_3=-\frac{22}{3}
\end{align}
\end{subequations}
solve the OP~\eqref{eq:OP}. According to Theorem~\ref{thm:sumEqualsMaxOutNeuron} the constraints of the OP guarantee \eqref{eq:sumEqualsMaxOutNeuron} for $\alpha_i=\hat{\alpha}_i$ and $\beta_i=\hat{\beta}$. Figure~\ref{fig:phi+h}(b) illustrates that in this case the modification of $\varphi(x)$ by the PWA function results in a symmetric function $\varphi(x)+h(x)$. 

According to Corollary~\ref{cor:ANNforPhi} a max-out-NN with one hidden layer of width $w_1=2$, $p_1=s=3$ and weights and biases as per \eqref{eq:WbforPhi} exactly describes the function \eqref{eq:example1_V1}. Since both solutions \eqref{eq:alphabetaAlg} and \eqref{eq:alphabetaOP} are such that \eqref{eq:conditionsLeftOfi} and \eqref{eq:conditionsRightOfi} hold, we can either use $\alpha_i=\tilde{\alpha}_i$ and $\beta_i=\tilde{\beta}_i$ or $\alpha_i=\hat{\alpha}_i$ and $\beta_i=\hat{\beta}_i$ for the weights and biases.

\subsection{2D PWQ function}\label{subsec:ex_2D}

\begin{figure}[ht]
\centering
\includegraphics[trim={2cm 11cm 3cm 11cm},clip,scale=0.5]{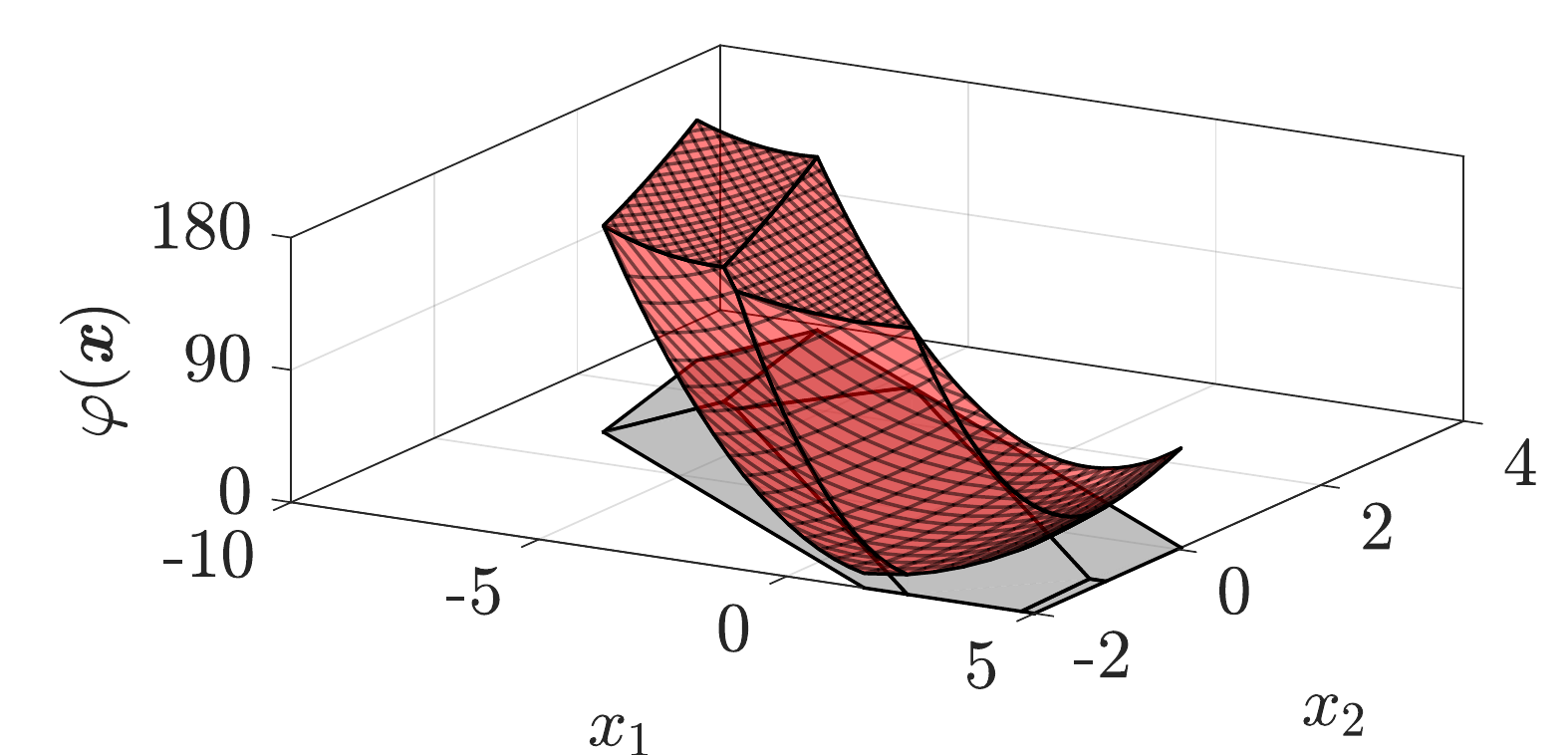}
\caption{Illustration of the convex PWQ function $\varphi(\xb)$ in red and the state space partition with 6 regions in gray.}
\label{fig:OVF2D}
\end{figure}

\begin{figure}[ht]
\centering
\includegraphics[trim={2cm 11cm 2cm 11cm},clip,scale=0.5]{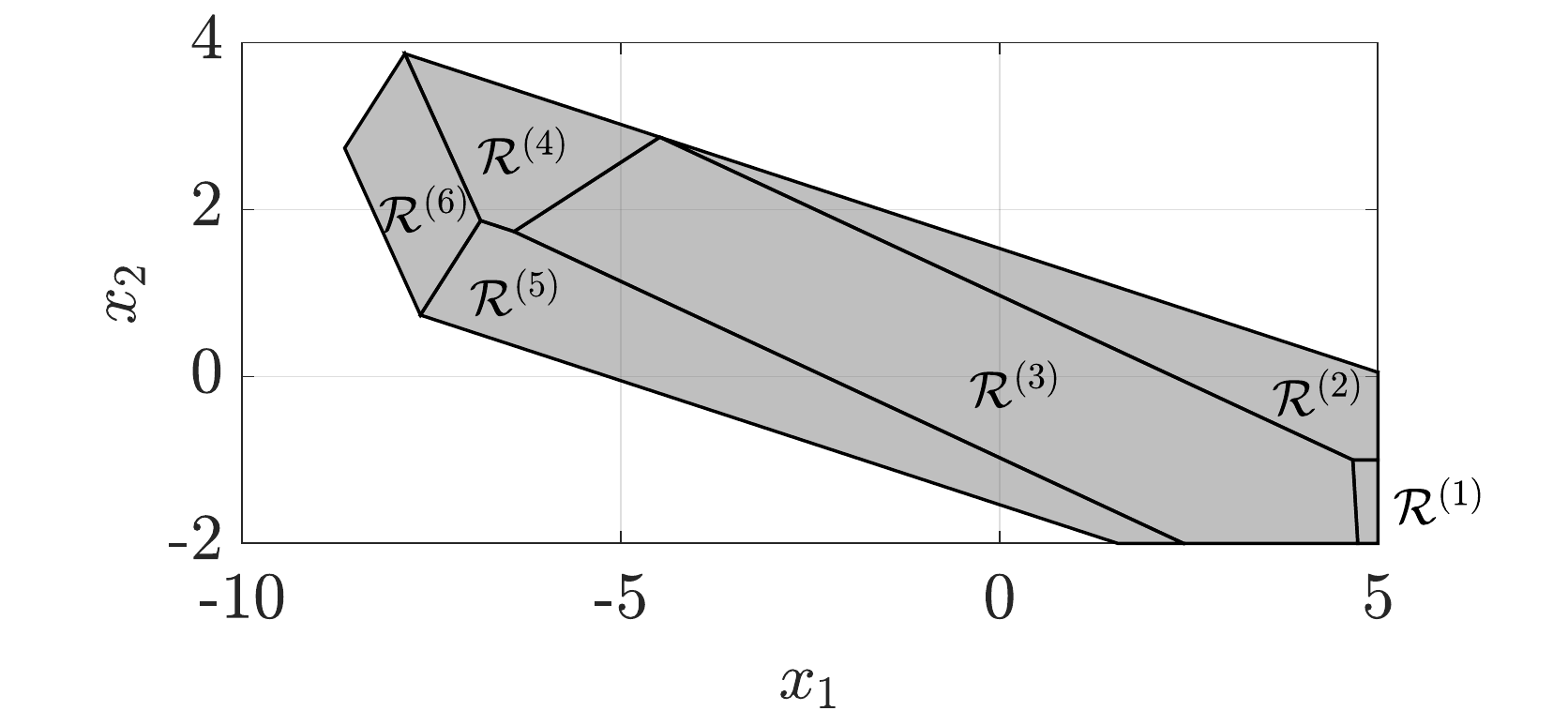}
\caption{Illustration of the partition of the functions $\varphi(\xb)$ and $h(\xb)$.}
\label{fig:partition2D}
\end{figure}

We consider the OCP
\begin{align}
\label{eq:OCP2D}
\varphi(\xb) := \!\!\!\!\min_{\substack{\xb_0,\xb^+,u}} 
\!\!\!\!\!\!\!\!\!\!\!\!\!\!\!\!\!\!\!\! & 
\,\,\,\,\,\,\,\,\,\,\,\,\,\,\,\,\,\,\,\,\,
\norm{\xb^+}_{\Pb}^2 +  \norm{\xb_0}_2^2 + u^2  \\
\nonumber
\text{s.t.}  \quad  \xb_0&=\xb, \\
\nonumber
 \xb^+&=\begin{pmatrix}1 & 1 \\0 & 1\end{pmatrix}\,\xb_0 + \begin{pmatrix}0.5 \\1\end{pmatrix} u, \\
 \nonumber
 \xb_0 & \in \{\xb\in\R^2 \ | \ -10 \leq x_1 \leq 5, \ -2 \leq x_1 \leq 4 \}, \\
 \nonumber
 \xb^+ & \in \{\xb\in\R^2 \ | \ \Gb \xb \leq \eb \},\\
 \nonumber
 u & \in \{u\in\R \ | \ |u| \leq 1\},\\[-6.5mm]
 \nonumber
\end{align}
with
\begin{align*}
\Gb:=\begin{pmatrix}
0 & -1\\
-0.43 & -1.03\\
0.43 & 1.03\\
0.43 & 0.03\\
-0.11 & 0.18
\end{pmatrix}, \quad
\eb:=\begin{pmatrix}
2\\
1\\
1\\
2\\
1
\end{pmatrix},
\end{align*}
which arises in the context of MPC for a system with double-integrator dynamics, where $\xb$ is the current system state, the prediction horizon is chosen as $N=1$ and where a stabilizing terminal set is considered. Explicitly solving~\eqref{eq:OCP2D} leads to the OVF shown in Figure~\ref{fig:OVF2D}, which is defined over six polyhedral regions as detailed in Figure~\ref{fig:partition2D}.

\vspace{-4mm}
Analogously to the first example, we also have 
\begin{equation*}
\varphi(\xb) \neq \max\limits_{1\leq i\leq 6}\left\{\varphi_i(\xb)\right\}
\end{equation*}
here. Nevertheless, it is possible to adapt the procedure used for the one-dimensional case by adding the convex PWA function
\begin{align*}
h(\xb)=\left\{
\begin{array}{cc}
99.2x_1  - 24.34 x_2 - 292.77 & \text{if}\,\,\,\xb\in\Rc^{(1)}, \\
99.2x_1 + 245.83 x_2 - 22.61 & \text{if}\,\,\,\xb\in\Rc^{(2)}, \\
-18.18x_1  - 32.03 x_2 + 247.56 & \text{if}\,\,\,\xb\in\Rc^{(3)}, \\
-24.21x_1  - 21.76 x_2 + 191.07 & \text{if}\,\,\,\xb\in\Rc^{(4)}, \\
-48.75x_1  - 104.4 x_2 + 177.19 & \text{if}\,\,\,\xb\in\Rc^{(5)}, \\
-107.26x_1  - 63.29 x_2 - 300.45 & \text{if}\,\,\,\xb\in\Rc^{(6)}, \\
\end{array}
\right.
\end{align*}
defined on the same partition (see Figure~\ref{fig:partition2D}) as the function $\varphi(\xb)$ which is such that the sum $\varphi(\xb)+h(\xb)$ can be represented in the form
\begin{equation}\label{eq:maxPhi+h_example2D}
\varphi(\xb)+h(\xb)=\max\limits_{1\leq i\leq 6}\left\{\varphi_i(\xb)+h_i(\xb)\right\}.
\end{equation}

With the relation~\eqref{eq:maxPhi+h_example2D} we can represent $\varphi(\xb)+h(\xb)$ in terms of a single max-out neuron. Which immediately allow to represent $\varphi(\xb)$ by a max-out-NN with one hidden layer and two neurons, as $h(\xb)$ itself can be represented by a second neuron (see \eqref{eq:maxPWA}). In contrast to the case of $n=1$, we need here an augmented input of the form
\begin{equation*}
\xib=\begin{pmatrix}
x_1 & x_2 & x_1^2 & x_1x_2 & x_2^2 \end{pmatrix}^\top,
\end{equation*}
because the max-out neuron for the representation $\varphi(\xb)+h(\xb)$ involves the evaluation of quadratic segments. Note that the presented exemplary solution for $n=2$ is in line with Conjecture~\ref{conj:maxoutPWQ}. 
\begin{figure}[ht]
\centering
\includegraphics[trim={2cm 11cm 3cm 11cm},clip,scale=0.5]{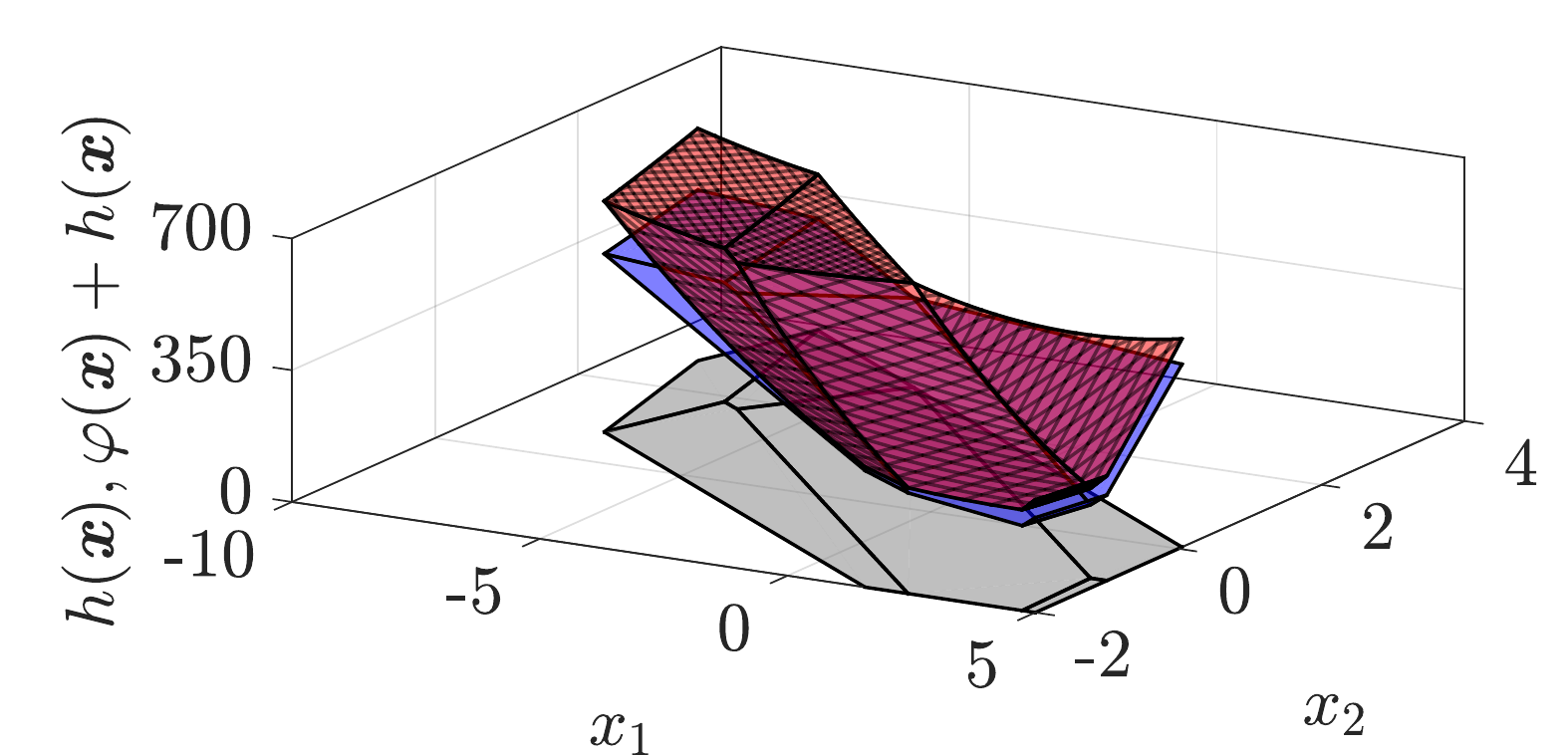}
\caption{Illustration of the lifted function $\varphi(\xb)+h(\xb)$ in red, the convex PWA function $h(\xb)$ in blue and the state space partition in gray. The sum $\varphi(\xb)+h(\xb)$ and the function $h(\xb)$ are represented by the first and second neuron of the max-out-NN, respectively.}
\label{fig:V+h}
\end{figure}

We briefly comment on the derivation of this solution. For the computation of the function $h(\xb)$ we first compute a convex lifting $H(\xb)$ of the state space partition of the function $\varphi(\xb)$ according to \cite[Alg.~3]{nguyen2017convex}. Then we choose $h(\xb)= \gamma H(\xb)$ with $\gamma>0$. The scaling factor $\gamma$ is chosen such that we have \eqref{eq:sumEqualsMaxOutNeuron}. Although this approach seems to work in general the proof that it is always possible to find a $\gamma$ such that \eqref{eq:sumEqualsMaxOutNeuron} holds and extending the conditions \eqref{eq:conditionsLeftOfi} and \eqref{eq:conditionsRightOfi} to $n>1$ is not straightforward. 

\section{Conclusions and outlook}\label{sec:conclusion}
In this paper, we proved that convex PWQ functions with scalar variables, i.e., $x\in\R$ can always be computed as the sum of two terms of the form \eqref{eq:maxPWA} and \eqref{eq:maxQuadPlusH}, which are the maximum of $s$ affine and $s$ quadratic segments, respectively (see Thm.~\ref{thm:sumEqualsMaxOutNeuron} and \ref{thm:feasibleSolution}).
We further showed in Corollary~\ref{cor:ANNforPhi} how to use this insight to derive a max-out-NN with one hidden layer of width $w_1=2$, which exactly represents the PWQ function. This result can be useful for deriving design guidelines for the choice of the topology when learning PWQ functions, as required in Q-learning for linear MPC. 

Example~\ref{subsec:ex_2D} shows that the central idea of adding a convex PWA function $h(\xb)$ to the original PWQ function such that \eqref{eq:maxQuadPlusH} holds is extendable to convex PWQ functions with $n>1$. The formal proof of this approach should involve the extension of the presented Theorems~\ref{thm:sumEqualsMaxOutNeuron} and \ref{thm:feasibleSolution} which is an interesting direction for future research.

% References

\end{document}